\documentclass[12pt]{article}
\usepackage{putex}
\usepackage{amsfonts}
\usepackage{amssymb}
\usepackage{amsmath}
\usepackage[bookmarks = false, colorlinks = true, linkcolor = blue, citecolor = purple]{hyperref}
\usepackage{ulem}
\usepackage{tipa}
\usepackage{graphics,graphicx}
\usepackage{setspace}
\numberwithin{equation}{section}

\newcommand{\HH}{\mathcal{H}}

\newcommand{\AAA}{\mathcal{A}}

\newcommand{\Tr}{\mbox{Tr}}

\newcommand{\OO}{\mathcal{O}}

\newcommand{\Sym}{\it Sym}
\newcommand\be{\begin{equation}}
\newcommand\ba{\begin{eqnarray}}
\newcommand\ee{\end{equation}}
\newcommand\ea{\end{eqnarray}}
\newcommand\bone{{\bf 1}}
\definecolor{purple}{rgb}{0.7,0.0,0.5}
\definecolor{huh}{rgb}{0.0,0.6,0.8}
\definecolor{orange}{rgb}{1,0.5,0}
\definecolor{pink}{rgb}{1,0.4,0.4}
\definecolor{light-gray}{gray}{0.75}

\begin{document}

\title{A Toy Model of Entwinement}
\authors{Jennifer Lin}
\institution{IAS}{{School of Natural Sciences, Institute for Advanced Study, Princeton, NJ, USA}}

\abstract{
Entwinement is the entanglement entropy of a subset of gauge-variant degrees of freedom in a certain twisted state of an orbifold CFT, defined by embedding the state in a larger Hilbert space with some gauge constraints removed. We propose an intrinsically gauge-invariant, algebraic definition of entwinement. Our main piece of evidence is a computation showing that, in a spin system that resembles the orbifold CFT, the analog of entwinement is the entanglement entropy of a gauge-invariant subalgebra, which we identify.
 We review why entwinement is relevant for the conjecture that entanglement builds spacetime.}
\maketitle

\newpage
 
 
\section{Introduction} 

Suppose that we are handed a holographic CFT with $R_{AdS} \gg \ell_s, \ell_P$. Then suppose that we are handed an excited state $|\Psi\rangle$ in the CFT. Questions that we'd like answered include: When does $|\Psi\rangle$ have a geometric dual description? If so, how can we read off the emergent bulk metric from data intrinsic to the CFT? And what low-energy CFT dynamics around $|\Psi\rangle$ give rise to the classical equations of motion (i.e. the Einstein equations) that constrain the bulk metric, at every point in the bulk?

These questions are closely related. For example, consider the third one. In order to understand what constrains the dynamics of the classical metric everywhere in the bulk, we need some idea how to define the metric from the CFT data. So the problem of deriving the Einstein equations around states of the CFT with geometric duals is morally equivalent to classical bulk reconstruction, and is a sharp litmus test for our understanding of emergent bulk locality, which is the real miracle of AdS/CFT. 


I will focus on this third problem. The reason is that while it is unsolved in general, one can cleanly solve it around empty AdS, and then see what is needed to generalize.

The linearized bulk Einstein equations (EFE's) around empty AdS were shown to be equivalent to the entanglement first law
\be\label{ee1stlaw}
\delta S_{EE} = \delta \langle H_{mod}\rangle
\ee
around the vacuum state of a dual CFT, considered for every ball-shaped spatial region $B$ in the CFT \cite{Lashkari:2013koa, Faulkner:2013ica, Swingle:2014uza}. Here $\delta S_{EE}$ is the change in the entanglement entropy (EE) across $B$ as one goes from the CFT vacuum to a perturbatively nearby state in the Hilbert space \footnote{I.e. ones where the orthogonal component to the vacuum is perturbatively small, $|\psi\rangle = (1-\epsilon)|0\rangle + \epsilon|\Psi\rangle$.}
, and $H_{mod} = -\log\rho$ is the modular Hamiltonian operator for the reduced density matrix $\rho$ of the ball-shaped region in the CFT vacuum, which happens to be a weighted integral of the CFT stress tensor over the ball. 

To briefly review the argument, one simply translates both sides of \eqref{ee1stlaw} to the bulk, assuming the relevant holographic dictionary entries as input. On the left-hand side one assumes the Ryu-Takayanagi formula \cite{Ryu:2006bv, Ryu:2006ef, Lewkowycz:2013nqa}
\be\label{rt}
S_{EE} = \frac{1}{4G_N} \min_{\partial A = \partial B}{\rm Area}(A)\,,
\ee
equating position-space entanglement entropies of the CFT to the minimal areas of homologous codimension-2 bulk surfaces. On the right-hand side, the change in the expectation value of the CFT stress tensor from its vacuum expectation value, and hence $\delta\langle H_{mod}\rangle$, is related to the leading asymptotic deviation of the bulk metric from pure AdS. In fact, this can also be shown to follow from the Ryu-Takayanagi formula \cite{Faulkner:2013ica}. 
Then eq. \eqref{ee1stlaw} turns into a bulk equation of the form
\be\label{eq1}
\int_{\partial A} F_1(\delta g_{\mu\nu}) = \int_B F_2(\delta g_{\mu\nu})
\ee
for each boundary ball $B$ with Ryu-Takayanagi surface $\partial A$. The family of equations \eqref{eq1} is in one-to-one correspondence with points in empty AdS, since for a given bulk point, there is a Ryu-Takayanagi surface for a ball-shaped region, whose apex passes through the point. This counting makes it possible that \eqref{eq1} can be inverted to yield the $tt$-component of the linearized EFE's at all points in the bulk, which turns out to be true. The other components of the EFE's follow from standard manipulations.

To summarize a bit more abstractly, first we found a collection of equations in the CFT to turn into the point-by-point EFE's. On the qualitative grounds that holography implements the RG, we anticipated a 1-parameter family of equations for each boundary point, with the extra parameter being an energy scale (above: the inverse radius of $B$). We then translated the equations to the bulk, assuming the relevant dictionary entries. Finally, we massaged the bulk expressions to turn them into the bulk EOM's. 
This argument demonstrates the equivalence of the assumed holographic dictionary entries and bulk classical gravity, suggesting that the assumed dictionary entries play a fundamental role in the organization of bulk locality. However, their origin must be independently explained. 

We'd like to generalize this logic to all asymptotically AdS spacetimes with a geometric bulk interpretation. 
To this end, the equations in the CFT that we want to turn to the bulk point-by-point EFE's must contain something whose bulk dual is a covariant functional of the metric, that has access to every point in the bulk, at least when the curvature can be made parametrically small.

Around vacuum AdS, the position-space EE served this purpose. However, in a generic asymptotically AdS spacetime without horizons, boundary-anchored minimal surfaces need not cover the bulk! It was experimentally discovered \cite{Freivogel:2014lja} that there can be regions of the bulk interior that no Ryu-Takayanagi surface passes through. This is sometimes called the {\it entanglement shadow}. To repeat our argument around general spacetimes, a necessary first step is thus to identify a shadowless bulk probe.

What types of covariant bulk probes are there in an asymptotically AdS spacetime? Given a boundary domain $B$ on a constant-time slice, two prototypical classes are surfaces defined by their causal relation to the domain (e.g. the surface of the bulk causal wedge anchored on the boundary causal diamond of $B$, and higher-codimension slices thereof \cite{Hubeny:2012wa}), as well as surfaces of extremal area anchored on $B$. The latter class contains the Ryu-Takayanagi surfaces of minimal area. The first class of surface is restricted to lie outside of bulk horizons by definition, and so cannot be used to reconstruct the bulk metric or derive the Einstein equations in some situations where we expect Einstein gravity to hold, e.g. in the eternal black hole at early times. As for the second class, we have seen that minimal-area surfaces are  experimentally insufficient for bulk reconstruction.

These considerations strongly motivate the following open question.

\begin{center}
{\it What is the CFT dual of non-minimal, extremal codimension-2 surfaces in the bulk?}	
\end{center}

Note that we want the answer in the strict classical limit. Non-minimal extremal surfaces are subleading corrections to the Ryu-Takayanagi formula at finite $N$ \cite{Lewkowycz:2013nqa}. This suggests that the infinite $N$ interpretation will have to do with entanglement.

The best hint that we have towards an answer comes from a class of examples of asymptotically AdS spacetimes, where non-minimal surfaces abound: $AdS_3$/$\mathbb{Z}_n$ conical defect geometries for integer $n$. By studying this example, the authors of \cite{Balasubramanian:2014sra} were led to conjecture that the area of a boundary-anchored, non-minimal, extremal codimension-2 bulk surface is dual to a novel gauge-invariant concept in the CFT that they called {\it entwinement}: the entanglement entropy of a certain subset of gauge-variant degrees of freedom in the homologous region of the CFT, defined by embedding the CFT state into a larger Hilbert space, with some gauge constraints lifted.

However, this definition is not wholly satisfying. It relies on the enlarged Hilbert space which is fictitious with respect to the physical CFT. A definition of entwinement intrinsic to the physical CFT would be nicer, perhaps shedding light on just how the CFT organizes into bulk geometry on sub-AdS scales.

In this note, we conjecture that the entwinement of \cite{Balasubramanian:2014sra} can be understood as the algebraic EE of a certain gauge-invariant operator subalgebra in the CFT, which is not the maximal subalgebra on a spatial region that enters the conventional notion of position-space EE and the Ryu-Takayanagi formula. We formally identify the subalgebra. Our main piece of evidence is an explicit demonstration of the conjecture in a simple spin model, that crudely resembles the relevant state in the orbifold CFT of \cite{Balasubramanian:2014sra}: a system of two spins with a discrete $\mathbb{Z}_2$ gauge symmetry, where both the analog of entwinement and the  EE of the  subalgebra can be computed and checked to agree. 

The plan of this note is as follows. In section \ref{s2}, I review the main result of \cite{Balasubramanian:2014sra}, explaining the origin of their conjecture that entwinement is the CFT dual of non-minimal extremal surfaces. In section \ref{s3}, I review the algebraic definition of entanglement entropy. In section \ref{s4}, I use the algebraic definition to show that the analog of entwinement in the $\mathbb{Z}_2$ spin system is the algebraic EE of a certain subalgebra (section \ref{s41}), and comment on the generalization to more complex systems (section \ref{s42}). I conclude in section \ref{s5}. The details of a computation are relegated to an appendix.

\section{Review of Entwinement} \label{s2} 

In this section, I review the origin of the conjectured relation between the areas of non-minimal, extremal surfaces and entwinement of the boundary CFT, following \cite{Balasubramanian:2014sra}. 

\begin{figure}
\centering
	\includegraphics[height=2in]{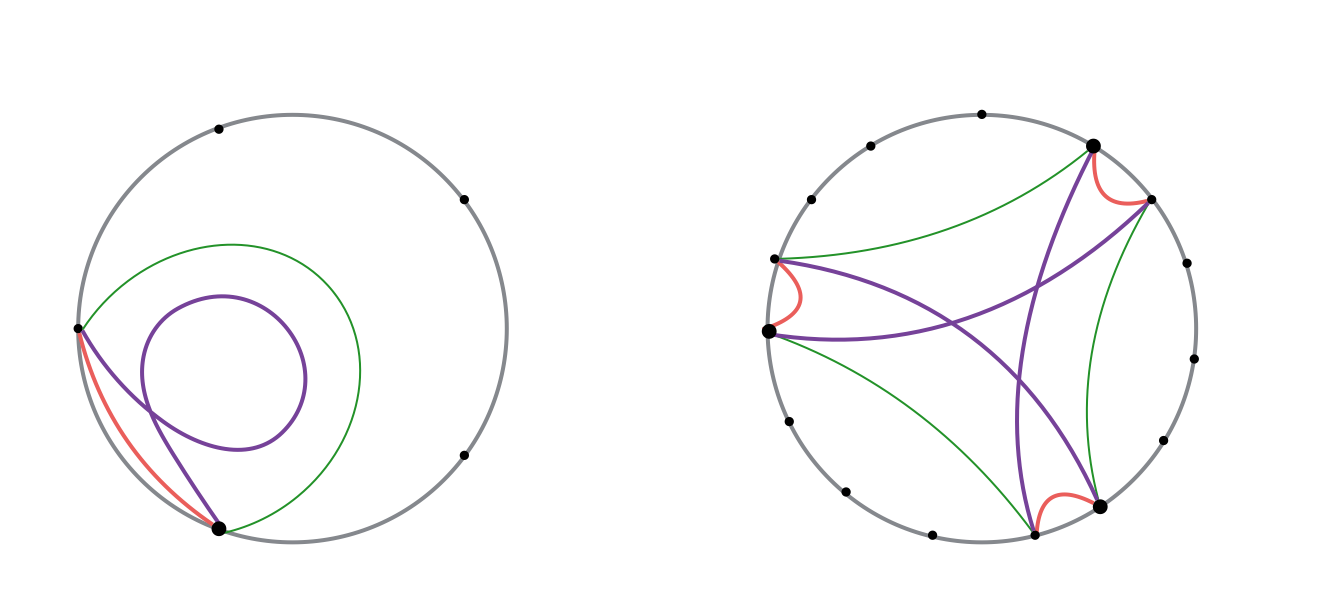}
	\caption{A constant-time slice of the $AdS_3/\mathbb{Z}_3$ conical defect geometry (left) and its covering space (right). The red line is a minimal-area codimension-2 surface (geodesic) in the defect background and the green and purple lines are non-minimal ones ending on the same boundary region. Their images in the covering space are shown. Anticipating later use, we have also split the boundary ${\bf S}^1$ into five regions. In the notation of section 4, this illustrates the example $(k,n) = (5,3)$. The features of the figure are  qualitatively correct, but it is not drawn to scale.}
	\label{fig:1}
\end{figure}

Consider the geometry $(AdS_3 \times {\bf S}^3)/\mathbb{Z}_n \times {\bf T}^4$ for integer $n$. The non-compact directions constitute the conical defect geometry AdS$_3/\mathbb{Z}_n$, which can be defined by taking global $AdS_3$ with coordinates
\be\label{metric}
ds^2 = -\left(1 + \frac{R^2}{L^2} \right)dT^2  + \left( 1+ \frac{R^2}{L^2} \right)^{-1} dR^2 + R^2 d\theta^2
\ee
and identifying the angular coordinate $\theta$ with period $2\pi/n$. Turning this around, the defect background has a covering space which is empty AdS$_3$ with a $n$-times-longer radius. We can compute geometric features of the defect background by performing the same computation over all its $\mathbb{Z}_n$ copies in the covering space. For example, suppose that we wanted to find the length of a minimal bulk curve homologous to a region of angular width $< 2\pi/n$ (in the coordinates \eqref{metric}) in the defect background. We can equivalently compute the lengths of the minimal bulk curves homologous to the $n$ copies of the region in the covering space, then divide by $n$. See Figure \ref{fig:1}.

From the covering space, it is easy to see that
\begin{itemize}
\item[(*)] The conical defect background has an entanglement shadow. I.e., there is a central region that no boundary-anchored minimal surface passes through. In the covering space, the shadow originates from the mutual information phase transition when we go from 
$n$ intervals of angular width $\alpha = \pi/n -\epsilon$ to $\alpha = \pi/n + \epsilon$.
\item[(*)] The conical defect background from quotienting by $\mathbb{Z}_n$ contains $n$ non-minimal geodesics homologous to each boundary interval on a constant-time slice. Moreover these non-minimal geodesics, together with the minimal ones, cover all the points in the bulk. We know this because the non-minimal geodesics are descended from minimal ones that subtend angles greater than $\pi/n$ in the covering space, which is just global AdS, with no entanglement shadow.
\end{itemize}

Regarding the defect background as an excited state of $AdS_3$,
 what is its CFT dual? Naively, since the defect is geometric, the dual is some state in a strongly coupled CFT, and there is not much more that we can say. Fortunately, we can do better by contemplating the defect background in the top-down D1-D5 brane construction. Here is a quick review of the construction. Consider $N_1$ D1 branes wrapping  ${\bf S}^1$ and $N_5$ D5 branes wrapping ${\bf S}^1 \times {\bf T}^4$. Defining $N = N_1N_5$, the near-horizon geometry is $AdS_3 \times {\bf S}^3 \times {\bf T^4}$ with $R_{AdS} \sim \sqrt{N}\,.$ By the usual logic, string theory on $AdS_3 \times {\bf S}^3 \times {\bf T^4}$ is dual to the low-energy limit of the worldvolume QFT, that here happens to be a 2d sigma model whose target space is a deformation of $({\bf T}^4)^{N}/\Sym(N)$. The permutation group $Sym(N)$ serves as the discrete nonabelian gauge group. The deformation parameter can be identified with one labeling a family of SUGRA solutions, giving a moduli space analogous to the $\lambda$ parameter labeling the moduli space of $N=4$ SYM. At the free orbifold point of the moduli space, analogous to $\lambda=0$ in $N=4$ SYM, the sigma model is a $N=(4,4)$ supersymmetric free field theory with target space $({\bf T}^4)^{N}/\Sym(N)$. The supergravity limit is at the far end of the moduli space. 

So we can ask: which state of the free orbifold CFT does the state dual to the $AdS_3/\mathbb{Z}_n$ conical defect interpolate to as we move along the D1-D5 moduli space? For $n$ a divisor of $N$, the answer turns out to be 
\be\label{twiststate}
\sigma^{N/n}|0\rangle
\ee
where $\sigma^{N/n}$ is the twist operator that sets up boundary conditions on the fields, s.t. they rotate into each other as 
\be\label{twistbc}
\begin{tabular}{ccccccccc}
$\phi_1$ & $\rightarrow $ & $\phi_2$ & $\rightarrow$ & $\dots$ & $\rightarrow$ & $\phi_n$  & $\rightarrow$ & $\phi_1$	\\
$\phi_{n+1}$ & $\rightarrow$ & $\phi_{n+2}$ & $\rightarrow$ & $\dots$ & $\rightarrow$ & $\phi_{2n}$ & $\rightarrow$ & $\phi_n$ \\
\vdots & & \vdots & & \vdots && \vdots && \vdots \\
$\phi_{N-n+1}$ & $\rightarrow$ & $\phi_{N-n+2}$ & $\rightarrow $ & $\dots$ & $\rightarrow$ & $\phi_{N}$ & $\rightarrow$ & $\phi_{N-n+1}$
\end{tabular}
\ee
as one goes around the spatial ${\bf S}^1$ of the CFT.  \footnote{Here and below we are keeping just one scalar from each $N=(4,4)$ superfield. For the fermions to be periodic too, there is a parity condition on $n$. We won't worry about this.} One can show this using string theory. Namely, one can explicitly construct the worldsheet theory of strings on $(AdS_3 \times {\bf S}^3)/\mathbb{Z}_n \times {\bf T}^4$ \cite{Martinec:2001cf}, and show that the background is BPS, hence protected as we interpolate in the moduli space.
The BPS states in the free orbifold CFT are the ground states in the twist sectors, labeled by conjugacy classes of the symmetric group. Finally, one can pick out the correct twist sector by U-dualizing the D1-D5 system to fundamental string frame, where the symmetric group representation of the oscillator excitations can be read off  \cite{Martinec:2002xq}.

The twisted state \eqref{twiststate} is related to the ground state in a theory of $N/n$ ${\bf T}^4$-valued scalar fields $\phi_{i=1, \dots, N/n}$, that are gauged under the permutation group $Sym(N/n)$ and live on an ${\bf S}^1$ $n$ times longer than the one we started with, in the same sense that the  conical defect geometry was related to the covering global $AdS_3$ in Figure \ref{fig:1}: the two states belong to completely different physical theories, but formally, computations in the twisted state of the original theory  give the same answers as the computation over all images in the ground state of the covering one \cite{Lunin:2000yv}. 

Both at the free orbifold point and at the semiclassical gravity end of the D1-D5 moduli space, a covering space construction naturally appears. Ref. \cite{Martinec:2002xq} conjectured that the covering space construction is protected along the D1-D5 moduli space. If true, we are led to the following CFT dual for non-minimal geodesics in the defect background. By an application of Ryu-Takayanagi in the covering space, the CFT dual of a non-minimal geodesic anchored to a boundary interval $B$ in the defect background is the EE of the degrees of freedom on one image of $B$ in the covering space. Since at weak coupling, the covering theory is obtained from the physical one by lifting some gauge constraints, one can heuristically say that the CFT dual of the non-minimal geodesic is the entanglement entropy of some ``gauge-variant" internal degrees of freedom in the CFT. It would be nice to have a more precise definition. This is our goal in the following sections.

In the rest of this note, I will forget about the bulk side of the entwinement story and just look for an intrinsic, gauge-invariant, definition of the entwinement in the state $\sigma^{N/n}|0\rangle$ of the free orbifold CFT: or indeed, in a simpler toy model thereof.

\section{Algebraic definition of entanglement entropy}\label{s3}

Entanglement entropy quantifies the ignorance of an observer who has access only to some of the observables in a quantum system. Conventionally, the entanglement entropies of QFT's are defined with respect to a spatial subregion of the QFT. In particular, this has been the definition relevant for the Ryu-Takayangi formula \eqref{rt} and for holographic applications so far. But EE can just as well be defined for any operator subalgebra in the QFT, that need not be associated with a subregion. We will argue that this more general concept has some relevance for holography. 

The algebraic definition of entanglement entropy has received some attention lately in the context of defining the EE in lattice gauge theories (see e.g. \cite{Donnelly:2011hn, Casini:2013rba, Radicevic:2014kqa, Ghosh:2015iwa, Radicevic:2015sza, Soni:2015yga, Radicevic:2016tlt}), though the goal there is to find a good definition of the EE of a subregion, in the presence of Hamiltonian constraints.

The definition is the following. Suppose that we are handed a quantum theory, specified by a Hilbert space $\HH$ and an algebra of observables. \footnote{It has been argued that in fact, algebraic EE can be defined without a priori specifying a Hilbert space \cite{Balachandran:2013cq}. Rather, given an operator algebra $\AAA$ and a state functional $\psi: \AAA\rightarrow \mathbb{C}$, a Hilbert space can be constructed using the GNS construction in algebraic QFT, and the EE computed with respect to it. However, we will always specify the Hilbert space in this note.} For completeness, we define an algebra to be a set of operators that is closed under addition, multiplication, and contains the identity element. 
 Given a state $\psi$ in $\HH$ and a subalgebra $\AAA_0$, we define the EE of the subalgebra to be the von Neumann entropy 
\be\label{defee}
S_{EE}(\AAA_0) = -\tr_{\HH}\rho \log \rho
\ee
of the unique element of the subalgebra, $\rho \in \AAA_0\,,$ s.t. 
\be\label{propa}
\tr_{\HH}(\rho \OO) = \langle\OO\rangle_{\psi}
\ee
for all operators $\OO$ in the subalgebra. 

Since $\rho$ is an element of $\AAA_0$, we can expand it as 
\be\label{defrho}
\rho = \sum_{\OO_i \in \AAA_0}\rho_i \OO_i\,,
\ee
where the $\rho_i$'s are c-numbers and index $i$ runs over linearly independent elements of $\AAA_0$. Condition \eqref{propa} gives one equation for each unknown $\rho_i$, so there is in general a unique solution. 

In this language, the EE of a subregion of a QFT, i.e. the von Neumann entropy of the reduced density matrix that one computes from taking a partial trace over part of the Hilbert space (assuming tensor product factorization), is the EE of the maximal subalgebra supported on the region.\footnote{Evaluated on the Hilbert space of the subregion. If we evaluate the EE of the maximal subalgebra WRT the global Hilbert space, there is an additive factor: see \eqref{cmp}.}  On the other hand, the algebraic definition is clearly more general, and has led to additional insights. For example, in a $U(1)$  gauge theory where the partial trace definition for position-space EE does not apply, non-distillable contributions to the position-space EE that appear in the algebraic definition were shown to be necessary to match the universal term of the EE across a sphere in four dimensions to the expected trace anomaly coefficient, resolving a longstanding puzzle \cite{Donnelly:2014fua, Huang:2014pfa}.   \footnote{It's worth noting that in {nonabelian} gauge theories, there is an interesting subtlety where the algebraic EE of the naively maximal gauge-invariant subalgebra on a spatial region appears not to give a good prescription for the position-space EE \cite{Soni:2015yga, Donnelly:2014gva}. For the definitions to agree, one should append certain edge operators to the algebra, such as Wilson lines ending on the entangling boundary \cite{Donnelly:2016auv}. The details of this story are well outside the scope of this paper.} The algebraic definition was also used recently to show how position-space EE is preserved across Ising dualities on the lattice \cite{Radicevic:2016tlt}, where a maximal algebra on a subregion generally dualizes to a non-maximal one. But perhaps most obviously and usefully, the algebraic definition allows us to assign an EE to any set of observables that we want, that need not have anything to do with spatial organization at all.

Algebraic EE is hard to compute. The prescription \eqref{defee}, \eqref{propa} may be well-defined, but it is usually intractable in practice, since one must enumerate all the operators in the subalgebra and solve a system of equations for each one. Of course, this is no surprise. The EE's of subregions from tracing out part of the Hilbert space are already famously difficult to compute in generic situations, and here we've passed to a generalization where we do not even have the benefit of the replica trick. 

However, the beautiful recent paper \cite{Radicevic:2016tlt} pointed out that in sufficiently simple spin models, algebraic EE's can in fact be computed quite explicitly. For a single spin with a 2d Hilbert space, the operator algebra consists of just the identity element and the Pauli matrices: $\AAA = \{\bone, \sigma_i\},\, i \in x, y, z$. (Below, we will sometimes use $\sigma$ to refer to all four of the operators $\{\bone, \sigma_i\}$). There are four subalgebras: $\{\bone\}, \{\bone, \sigma_x\}, \{\bone, \sigma_y\}, \{\bone, \sigma_z\}\,$. 
	From the orthogonality of the Pauli matrices, $\Tr(\sigma^a\sigma^b) = 2\delta^{ab}$, one can easily show that in any state of a system made from tensoring the Hilbert spaces of single spins, upon expanding the global density matrix $\rho$ in the form \eqref{defrho},
	 the coefficients $\rho_i$ are  
	\be\label{spinrho}
	\rho_i = \frac{\langle\OO_i\rangle}{\dim\HH}\,.
	\ee
Moreover, if we know the $\rho_i$'s for the global density matrix, we can get the reduced density matrix of any subalgebra by  projecting the expansion \eqref{defrho} onto the elements of the subalgebra. This again follows from orthogonality. 
	Spin systems are thus an especially nice playground in which to begin exploring the features of algebraic EE.
		
In the rest of this note, we will exploit the simplicity of spin systems to demonstrate that in discrete toy models of the symmetric orbifold CFT described in Section \ref{s2}, the analog of entwinement -- the EE of a subset of ``gauge-variant" degrees of freedom in a region,  defined by embedding the state into a larger Hilbert space where the gauge constraints are removed -- can be formulated as the algebraic EE of a non-maximal subalgebra in the region.

\section{Entwinement in spin systems}\label{s4}

In the free orbifold CFT of section \ref{s2}, let us set $N=n$ for simplicity. We put the CFT on a lattice, discretizing the spatial ${\bf S}^1$ into $k$ cells of angular size $2\pi/k$, and replacing the $n$ scalar fields with $n k$ quantum variables $s_i^m,\,\,i \in 1, \dots, n,\,\, m \in 1, \dots, k$. Moreover, we take the $s$'s to be $\mathbb{Z}_2$-valued instead of ${\bf T}^4$-valued: i.e., they are now quantum spins.

We have now a system of $nk$ spins, with $n$ spins on each of $k$ lattice sites. The discrete gauge group $Sym(n)$ symmetrizes the spins at a given site. The gauge condition projects out a part of the Hilbert space relative to the naive $2^{nk}$-dimensional tensor product Hilbert space. The boundary of the LHS of figure \ref{fig:1}, with each site carrying three spins, illustrates our discrete toy model with $(k=5, n=3)$. 

This model is a cousin of the orbifold CFT that entered the story of section \ref{s2}. On the other hand, for the reasons explained in section \ref{s3}, we will be able to do explicit computations of the algebraic EE's of various subalgebras. 

In a 2d QFT on ${\bf S}^1 \times \mathbb{R}$, the spatial ${\bf S}^1$ puts a physical constraint on the algebra of observables, which is that fields should satisfy certain boundary conditions as one goes around it. In particular, for orbifold theories the Hilbert space is a direct sum over twist sectors, each distinguished by a different boundary condition as one goes around the ${\bf S}^1$. Moreover, for the purpose of comparing to the example of Section \ref{s2}, we care about one particular set of boundary conditions, \eqref{twistbc}. 
 
In the toy model, this information could be crudely captured by imposing constraints that relate some $s_k^i$'s and $s_1^{i'}$'s. How to more precisely implement twist operators in such models is a separate interesting question. Rather than worrying about this for now, we will make comments that hold for all states in the spin system, including whichever one best resembles a discretization of \eqref{twiststate}. The main feature that we want to study is not the boundary conditions, but what happens when we un-gauge the system by a subgroup $\mathbb{Z}_n \in \Sym(n)$. 
     
\subsection{Entwinement in a two-spin system with $\mathbb{Z}_2$ gauge symmetry}\label{s41}

Consider a system of two spins with a $\mathbb{Z}_2$ gauge symmetry acting on the spins, allowing only states that are invariant under spin exchange. In the immediately preceding notation, this is the case $(k=1, n=2)$. The generalization of what we write here to $n>2$ with gauge group $Sym(n)$ is straightforward; the generalization to $k>1$ will be discussed later. 

Starting from the 4d Hilbert space for the tensor product of two spins, we project out the singlet state $\frac{1}{\sqrt{2}}(|\uparrow\downarrow\rangle - |\downarrow\uparrow\rangle)$. The Gauss law constraint implementing this is 
\be\label{proj}
S^2 |\psi\rangle = |\psi\rangle
\ee
where $S^2$ is the total spin operator 
\be\label{gauss}
S^2 = S_1^2 + S_2^2 + 2 {\bf S}_1 \otimes {\bf S}_2 = \frac 34 \bone \otimes \bone + \frac 14 (\sigma_x \otimes \sigma_x + \sigma_y \otimes \sigma_y + \sigma_z \otimes \sigma_z)\,
\ee
and the vector ${\bf S}$ is $(\sigma_x, \sigma_y, \sigma_z)$.
$S^2$ also doubles as the projection operator onto gauge-invariant states, since it has eigenvalue 1 on states invariant under spin exchange and eigenvalue 0 on the singlet state. 

Of the sixteen linearly independent operators $\sigma \otimes \sigma$ that constitute the algebra in the theory without gauge constraints, nine combinations of them commute with $S^2$ and are gauge-invariant. They are 
\be\label{z2alg}
\AAA: \{\bone \otimes \bone\,,\,\, \sigma_i \otimes \sigma_i\,,\,\, \frac 12(\bone \otimes \sigma_i + \sigma_i \otimes \bone),\,\, \frac 12 \sigma_{(i} \otimes \sigma_{j)}\}
\ee
where $i$ runs over $\{x, y, z\}$ and $i \neq j$. List \eqref{z2alg} contains ten operators, but the Gauss law trivializes one linear combination of them. It is quite obvious by symmetry that the operators in list \eqref{z2alg} are the ones that are allowed under the spin exchange.

All of the operators in \eqref{z2alg} have the property $\OO^3 = \OO$. 
 Hence, subalgebras of $\AAA$ include 
\be\label{z2salg}
 \{\bone \otimes \bone\,,\,\, \OO\,,\,\, \OO^2\} \qquad \forall\,\, \OO \in \AAA\,.
\ee

We will now show that in any state of this system, the analog of entwinement, which is the entanglement entropy between the two spins when the state is lifted to the 4d Hilbert space without the $\mathbb{Z}_2$ gauge constraint,
 is equal to the algebraic EE of a particular subalgebra of the form \eqref{z2salg}. The subalgebra depends on the state. Again we emphasize that in defining ``entwinement" for any state of the spin system, we are overreaching the analogy with the continuum example in section \ref{s2}, which is concerned specifically with the state \eqref{twiststate}. We will come back to this point later.

Below we will call the 4d Hilbert space the ``extended Hilbert space" and sometimes denote it as $\HH_{ext.}$ In the context of taking the partial trace in the extended Hilbert space, we will sometimes label the two spins as $V$ and $\bar{V}$ and their 2d Hilbert spaces as $\HH_V$ and  $\HH_{\bar V}$.

We will first write down a formula for the entwinement in any state $\psi$ of this system, without referencing the extended Hilbert space, then show that it is the algebraic EE of a subalgebra. Given any state of the two-spin system, let $p_1$ and $p_3$ be the coefficients of the operators $\frac 12[\bone \otimes \sigma_x + \sigma_x \otimes \bone]$ and $\frac 12[\bone \otimes \sigma_z + \sigma_z \otimes \bone]$, respectively, when we expand the global density matrix in the form \eqref{defrho}. Then the entwinement is
\be\label{entform}
S_{ent.}(\psi) = -\left[\frac 12 - \sqrt{p_1^2+p_3^2} \right]\log\left[\frac 12 - \sqrt{p_1^2+p_3^2} \right] - \left[\frac 12 + \sqrt{p_1^2+p_3^2} \right] \log \left[\frac 12 + \sqrt{p_1^2+p_3^2} \right]\,.
\ee
 
To derive \eqref{entform}, we work backwards from the partial trace in the extended Hilbert space. The steps are the following.

\begin{enumerate}
\item {\it EE from the partial trace vs. algebraic EE in the extended Hilbert space}. \\
First, note that the usual EE from the von Neumann entropy of the density matrix obtained by taking the partial trace over one spin $\bar{V}$, is related to the algebraic EE of the maximal subalgebra supported on the other spin by an additive constant,
\be\label{cmp}
S_{EE}(\mbox{max subalgebra on V}) = S_{EE}(\mbox{partial trace over ${\bar V}$}) - \log\frac{\dim\HH_{V}}{\dim\HH}\,.
\ee

Eq. \eqref{cmp} holds for any system where we want to compare the EE's from partial trace and from the maximal subalgebra.
The constant appears because after we take a partial trace, the reduced density matrix $\rho_V$ is thought as acting on the smaller Hilbert space $\HH_V$, while in the algebraic approach we identify a density operator that is defined on the full Hilbert space $\HH_V \otimes \HH_{\bar V}$. 
For the two-spin system, $\frac{\dim\HH_V}{\dim\HH_{ext.}} = \frac 12$. 
\\
Note that the constant $-\log \frac{\dim \HH_V}{\dim\HH} = \log\dim\HH_{\bar V}$ is a positive number, so the EE from the maximal subalgebra is greater than the EE from the partial trace.

%

\item {\it Embed the state $\psi$ in the extended Hilbert space, and compute the reduced density matrix for the maximal subalgebra on one spin there}. 

Every state in the gauge theory can be uniquely embedded into the extended Hilbert space with the gauge constraint removed. We do this. Then we read off the projection of the global density matrix \eqref{defrho} onto those operators that act nontrivially on just one spin. As explained around \eqref{spinrho}, this projection gives the reduced density matrix for the maximal subalgebra supported on that spin.

  The answer turns out to be 
\be\label{redrho}
\rho_{\AAA(V)} = \frac 12 P \otimes \bone \ee
where
\be \label{defp} \qquad P = \frac 12 \bone +  p_1 \sigma_1 +  p_3 \sigma_3\,.
\ee
It's a function of the state that we started with through the coefficients $p_1$ and $p_3$. 


\item {\it Simplify the expression for the von Neumann entropy of the reduced density matrix \eqref{redrho} in the extended Hilbert space}.
%
%
At this point, the entwinement is  
\be\label{defentw}
S_{ent.}  + \log 2 = \frac 12\tr_{\HH_{ext.}}\left[\left(\frac 12P \otimes \bone\right) \log \left(\frac 12P \otimes \bone\right) + \left(\bone \otimes \frac 12P\right) \log \left(\bone \otimes \frac 12P\right)\right]
\ee
where the RHS is the von Neumann entropy of the density matrix \eqref{redrho}, and  
the $\log 2$ on the LHS comes from \eqref{cmp}\,. We would like to simplify the RHS. Plugging \eqref{defentw} into the replica trick formula and massaging the equations, with $p = P/2$, we find 
\begin{eqnarray}
	2S_{ent.} + 2 \log 2 &=& -\partial_n\left(\log\tr_{\HH_{ext.}}(p \otimes \bone)^n + \log \tr_{\HH_{ext.}}(\bone \otimes p)^n\right)|_{n \rightarrow 1} \\
	&=& -\partial_n \log\tr_{\HH_{ext.}^2}(p \otimes \bone)^n \otimes (\bone \otimes p)^n|_{n \rightarrow 1} \\ 
	&=& -\partial_n\log\tr_{\HH_V^2}(p \otimes p)^n|_{n \rightarrow 1}\\
	&=& -\tr_{\HH_V^2} (P \otimes P) \log (P \otimes P) \\
	&=& -2\tr_{\HH_V} P \log P\,\,.  \label{lastline}
\end{eqnarray}
Here $\HH_{ext.}^2$ is the 16d Hilbert space of four spins  in the natural basis for the Kronecker product, and $\HH_V^2$ is the Hilbert space of two spins. Note that in the second-to-last line, we picked up some factors in reversing the replica trick formula because  $\tr p \neq 1$. 

\item {\it  } Finally, to obtain \eqref{entform}, we took the exact form of $P$ as a function of $p_1$ and $p_3$ \eqref{redrho} and explicitly computed its von Neumann entropy in the 2d Hilbert space. 

\end{enumerate}

 This completes the explanation of eq. \eqref{entform}. 
 Now we will show that, for any state, \eqref{entform} is the algebraic entanglement entropy \eqref{defee}, \eqref{propa} of a subalgebra of \eqref{z2alg}.
 
  In fact, if we hypothesize that \eqref{entform} is the EE of some subalgebra, there is only one thing that it can be. Since the coefficients $p_1$ and $p_3$ appear explicitly in \eqref{entform}, the subalgebra must involve the operators $\frac 12[\bone \otimes \sigma_x + \sigma_x \otimes \bone]$ and $\frac 12[\bone\otimes \sigma_z + \sigma_z \otimes \bone]$, that carry those coefficients. On the other hand, the algebra generated by both these operators can easily be checked to be too large: the reduced density matrix for it will be the full density matrix in any state. This leaves the natural guess that the desired subalgebra is of the form \eqref{z2salg}, for  
\be\label{defo}
\OO = p_1 \cdot \frac 12[\bone \otimes \sigma_x + \sigma_x \otimes \bone] + p_3 \cdot \frac 12[\bone\otimes \sigma_z + \sigma_z \otimes \bone].
\ee
The operator $\OO$ depends on the state through the coefficients $p_1$ and $p_3$. So the entwinement in this model is the algebraic EE of a state-dependent subalgebra.

It remains to be shown that the guess is correct. In the appendix, we explicitly compute the algebraic EE of the subalgebra generated by \eqref{defo} and check the guess by brute force.

To summarize, we found that in the two-spin system with a $\mathbb{Z}_2$ gauge symmetry, the entwinement in any state of the system, that we defined to be the entanglement between the spins when the state is embedded in an extended Hilbert space with the $\mathbb{Z}_2$ gauge constraint lifted, is the algebraic EE of the subalgebra generated by the projection of the global state density matrix onto operators of the form  $\bone \otimes \sigma_i + \sigma_i \otimes \bone$, \eqref{defo}. \footnote{$\bone \otimes\sigma_y + \sigma_y \otimes \bone$ never contributes because its matrix representation is antisymmetric.}

This model of two spins with $\mathbb{Z}_2$ gauge symmetry is a trivial case in the discretization of the orbifold CFT described at the start of this section, where we take the entire spatial ${\bf S}^1$ to be a single lattice site. So the entwinement in this model does not have a geometric interpretation when we replace the spins by discrete field variables, take the continuum limit, and interpolate back to strong coupling.

\subsection{Generalizing to $k>1$ and the continuum}\label{s42}

In order to have a notion of locality, we must increase the parameter $k$ that counts the number of lattice sites in the discretization of the ${\bf S}^1$. So suppose that we have not two but $2k$ spins, with a $\mathbb{Z}_2$ gauge symmetry exchanging one group of $k$ with the other. The spins $s_i^m$ are labeled by the two indices $i \in 1,2$, and $m \in 1, \dots, k$. The extended Hilbert space $\HH_{ext.}$ is the $2^{2k}$-dimensional tensor product of the Hilbert spaces for the individual spins. 

Consider the subset of spins with index $m = 1, \dots, k'$ for $k' < k$. In the continuum limit, these are all the degrees of freedom in a region of angular width $2\pi k/k$ on the spatial ${\bf S}^1$. In the state \eqref{twiststate} of the continuum CFT, the associated boundary region of the dual theory at strong coupling has two bulk geodesics ending on it, one minimal and one that encircles the conical defect. The lengths of the two geodesics correspond to the strong coupling continuation of the entwinement of the spins $\{s_1^m\}$ with $m = 1, \dots, k'$ and with $m = k'+1, \dots, k$. 

For an arbitrary state in the spin model, we again define the entwinement of a subset of gauge-variant spins to be the EE of those spins when the state is embedded in an extended Hilbert space, with the $\mathbb{Z}_2$ gauge constraint lifted.

\begin{figure}
\centering
	\includegraphics[height=2in]{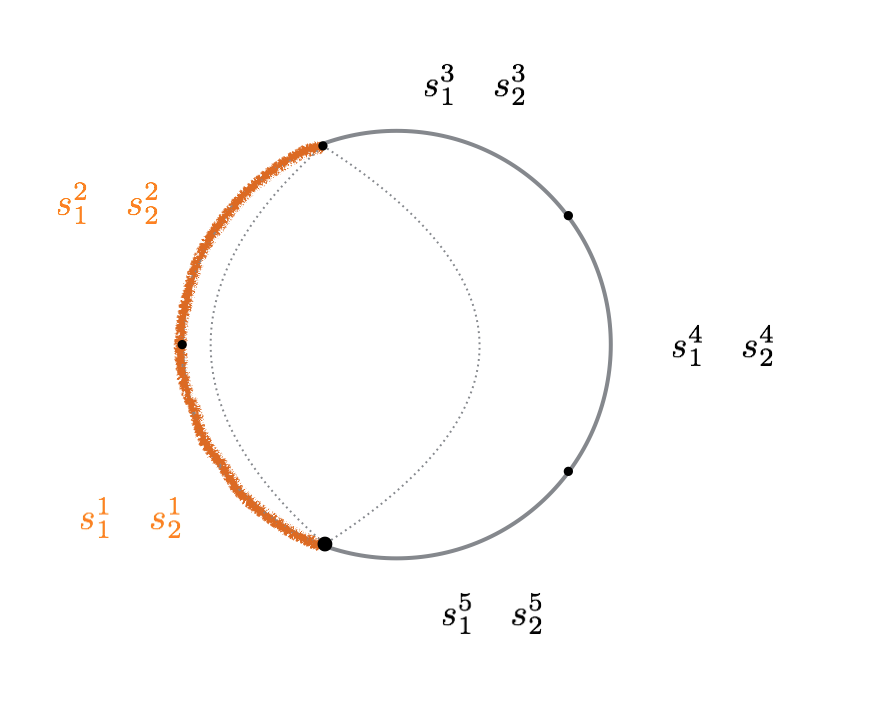}
	\caption{The ten-spin system for $k=5, n=2$ with the geometric interpretation illustrated. The orange spins form the subset $k'=2$. The associated region and  geodesics are drawn (not to scale).}
	\label{fig:2}
\end{figure}

Is this entwinement, say for the subset of spins $V: \{s_1^m\},\,\,m = 1, \dots, k'$, also the algebraic EE of a gauge-invariant subalgebra? We will not answer this question explicitly but will make a plausibility argument. Let us look to the previous $k=1$ case, and see which elements of that story generalize. In the steps leading up to \eqref{entform}, a version of \eqref{lastline} holds for all $k>1$ up to changing the additive constant on the LHS from $2 \log 2$ to $2 \log \dim\HH_{\bar V} = 2(k-k')\log 2$. But the operator $P$ \eqref{redrho}, rather than being a function of just the two coefficients $p_1$ and $p_3$, will receive contributions from all operators of the form 
\be\label{opf}
\OO_V \otimes \bone_{\bar V} + \mbox{$\mathbb{Z}_2$-symmetrized copy}
\ee
in the theory, with a linear dependence on their state-dependent coefficients in the expansion \eqref{defrho} of the global density matrix.

If the entwinement is to be the EE of a subalgebra then, there seems to again be a unique choice of the subalgebra. On the one hand, the coefficients of all operators of the form  \eqref{opf}, in the expansion \eqref{defrho} of the density matrix, must appear. On the other hand, the subalgebra generated by all such operators is the maximal position-space subalgebra. To keep the dependence on all their coefficients in \eqref{defrho} without taking the maximal subalgebra, we project the global density matrix onto operators of the form \eqref{opf}, call the projected operator $\OO$, and conjecture that the subalgebra defining entwinement is the one generated by $\OO$.

We have not checked this beyond $k=1$, but the following comments are circumstantial evidence:
\begin{enumerate}
\item It is true for $k=1$.
\item We argued that if entwinement in the spin model is an algebraic EE, it is what we just said based on the structure of the considerations following eq. \eqref{entform}. It is not necessary that entwinement is an algebraic EE. However, as algebraic EE is the most general gauge-invariant notion of entanglement, it is hard to imagine what other type of quantity it can be. 
\item A generalization of our claim is known to be true in other situations. Namely, we claim that the EE of a subset of degrees of freedom $V$ when a state of an abelian gauge theory is embedded in a larger Hilbert space with the gauge constraints removed, is the algebraic EE of the subalgebra generated by the projection of that state's density matrix onto operators of the form $(\OO_V \otimes \bone_{\bar V} + $ gauge copies). This is known to be true in lattice gauge theories, where it is the statement that the position-space EE in the extended Hilbert space of \cite{Ghosh:2015iwa, Soni:2015yga} is equal to the algebraic EE in the electric center choice of \cite{Casini:2013rba}. (In this context, our ``projection" step is a convoluted, implicit way of defining the electric center).
\end{enumerate}

Taking $k \rightarrow \infty$, we recover the continuum.

To summarize, we conjecture that in the AdS$_3$/CFT$_2$ example of \cite{Balasubramanian:2014sra}, the CFT duals to geodesics anchored on a boundary interval in the conical defect background are the algebraic EE's of the subalgebra generated by the operator $\OO$, that one gets from projecting the global density matrix of the  state \eqref{twiststate} onto those gauge-invariant operators of the form \eqref{opf}, for the relevant subset of degrees of freedom $V$ (i.e. the region subtended by one image of the geodesic in the cover, see Figure \ref{fig:1}). 

This definition is somewhat formal, with the generator of the subalgebra defined by the projection of the density matrix onto a certain subset of operators. It would be nice to more explicitly identify the relevant subalgebra for the twisted state \eqref{twiststate}. This might allow some consistency checks. E.g. for sufficiently small regions, the entwinement of \cite{Balasubramanian:2014sra} should reduce to the position-space EE at leading order in $1/N$ (this fact is obtained from the bulk geometry). This would be nice to check. We leave this issue for the future.



\section{Discussion} \label{s5}

Does spacetime emerge from quantum entanglement \cite{VanRaamsdonk:2009ar, VanRaamsdonk:2010pw, Bianchi:2012ev, Maldacena:2013xja,  Jacobson:2015hqa}? This hypothesis takes on its sharpest form in AdS/CFT, where the Ryu-Takayanagi formula \cite{Ryu:2006bv, Ryu:2006ef} and its generalizations \cite{Hubeny:2007xt, Faulkner:2013ana} have led to the nascent establishment of a ``holographic entanglement dictionary," including the celebrated derivation of the linearized Einstein equations from entanglement dynamics \cite{Lashkari:2013koa, Faulkner:2013ica, Swingle:2014uza}. 
However, the canonical notion of position-space EE in the CFT cannot be the end of the story. This is because the dual boundary-anchored minimal surfaces of the Ryu-Takayanagi formula and HRT generalization have entanglement shadows \cite{Freivogel:2014lja}, making them impotent for bulk reconstruction across all semiclassical asymptotically AdS geometries. \footnote{If indeed ``entanglement builds spacetime" in holography, it is also obvious that the entanglement of non-maximal algebras in the CFT must be considered to reconstruct the bulk in $(0+1)d/(1+1)d$ examples.} 

The main purpose of this note was to draw attention to the relevance for holography of EE's of subalgebras not organized by position space. In particular, we argued that the entwinement of \cite{Balasubramanian:2014sra} -- the EE of gauge-variant degrees of freedom, defined by embedding a state in an extended Hilbert space with the gauge constraints removed -- has an equivalent definition as the algebraic EE of a certain non-maximal, gauge-invariant operator subalgebra. Our main evidence was the calculation in section \ref{s41} showing that this statement is true for all states in a simple spin model, that crudely resembles the orbifold CFT studied in \cite{Balasubramanian:2014sra}, and comments in section \ref{s42} on the generalization to more complicated models. 

I conclude with some follow-up directions.

\subsubsection*{\it CFT dual of non-minimal bulk surfaces}
 
The entwinement in \cite{Balasubramanian:2014sra} was argued to be dual to the areas of non-minimal codimension-2 bulk extremal surfaces, for a specific class of states in orbifold CFT's. The most pressing direction for follow-up work is to understand when and how this idea generalizes. Any subalgebra of a CFT has an associated, well-defined EE \eqref{defee}, in any state; which of these have geometric interpretations? Do the EE's of certain subalgebras define the areas of bulk surfaces? One can define an RG flow for states by picking out an operator subalgebra and projecting the density matrix \eqref{defrho} onto it at each coarse-graining step. For a privileged choice of algebras, is this how holography implements the RG?

In trying to make these ideas precise, new technology is needed. For example, as explained earlier, no machinery exists to compute algebraic EE's in QFT's (or indeed, in quantum theories more complicated than spin chains, where we heavily utilized the Pauli matrix algebra). Perhaps we can develop a version of the replica trick that applies to certain non-maximal algebras on subregions, in order to someday connect with the Euclidean gravity approach of \cite{Lewkowycz:2013nqa}.
 For preliminary work in this direction, see e.g. the literature on EE in lattice gauge theory and \cite{Ohmori:2014eia} in the continuum for discussions about the replica trick when one removes some operators from the entangling edge.
 
In the meantime, we might try to generalize from the orbifold CFT to more conventional CFT's with continuum gauge groups, but it is not obvious to me how to do this. The analogy between orbifold CFT's and CFT's with continuum gauge groups is cleanest on worldsheets \cite{Pakman:2009zz}, but the Ryu-Takayanagi formula is not well-understood from the worldsheet point of view; it is an open problem how to define a quantity in string theory that resums the $\alpha'$  corrections to the Ryu-Takayanagi formula.

\subsubsection*{\it Generalizing the vacuum derivation of the linearized EFE's}

If indeed the areas of non-minimal extremal surfaces can be shown to be dual to certain algebraic EE's, it opens the way to extend the demonstrated equivalence of the linearized Einstein equations and the entanglement first law \eqref{ee1stlaw} in vacuum AdS, to generic spacetimes, with entanglement shadows. Conversely, if the slogan ``entanglement dynamics = gravitational dynamics" is to be true in general, some version of this must be true.\footnote{E.g., if the story in \cite{Jacobson:2015hqa} is correct we must be able to pull it onto the AdS boundary.}

In order to complete such a program though, in addition to the CFT dual of non-minimal surfaces, we would also need to define the gravity dual of modular Hamiltonians for non-maximal algebras on subregions. 

\subsubsection*{\it Algebraic EE and the physical meaning of the Ryu-Takayanagi formula}

If a canonical understanding of the RT formula can be acquired (i.e. what is the area term counting), it is easy to imagine that the lessons learned would have implications for the CFT duals of other bulk surfaces, in addition to being tremendously interesting in its own right. The recent paper \cite{Harlow:2016vwg}, which also emphasized the relevance of algebraic EE for holography, pointed out that assuming AdS/CFT is a quantum error correcting code \cite{Almheiri:2014lwa}, for states in the code subspace, the entanglement between a subregion of the CFT and its complement equals the algebraic EE of the maximal subalgebra supported in the bulk entanglement wedge, plus the expectation value of a center operator to be identified with the Ryu-Takayanagi area. Comparing this to the formula for EE in nonabelian lattice gauge theories (where in addition to distillable quantum EE one finds various edge terms) \cite{Donnelly:2014gva}, one is led to identify the Ryu-Takayanagi area with a particular edge term counting the dimension of gauge group representations on lattice links cut by the entangling surface. This observation was also made in \cite{Donnelly:2016auv}. 


We leave the exploration of these exciting ideas to the future.

\subsection*{Acknowledgments}
I am grateful to Daniel Harlow, Aitor Lewkowycz, and especially Djordje Radicevic for discussions. My work is supported by the Schmidt Fellowship and the U.S. Department of Energy.

\begin{appendix}
\section{Brute-force verification of the result of section \ref{s41}}	

In this section we prove that for any state in the system of two spins with $\mathbb{Z}_2$ gauge symmetry, the entwinement \eqref{entform} is the algebraic entanglement entropy of the subalgebra generated by the operator 
\be\label{defoo}
\OO = p_1\cdot\frac 12(\bone \otimes \sigma_x + \sigma_x \otimes \bone) + p_3 \cdot \frac 12(\bone \otimes \sigma_z + \sigma_z \otimes \bone)
\ee
where $p_1$ and $p_3$ are the coefficients for the operators $\frac 12(\bone \otimes \sigma_x + \sigma_x \otimes \bone)$ and $\frac 12(\bone \otimes \sigma_z + \sigma_z \otimes \bone)
$ in the expansion \eqref{defrho} of the global state density matrix. Hence the choice of the subalgebra is state-dependent.

Consider the most general pure state in the $\mathbb{Z}_2$ spin system 
\be \label{state}
a|\uparrow\uparrow\rangle + b|\downarrow\downarrow\rangle + \frac{c}{\sqrt{2}}(|\uparrow\downarrow\rangle + |\downarrow\uparrow\rangle), \qquad a^2+b^2+c^2=1\,.
\ee
By explicitly computing the matrix representations of the operators \eqref{z2alg} in the basis $|\uparrow\uparrow\rangle$, $|\downarrow\downarrow\rangle$, $\frac{1}{\sqrt{2}}(|\uparrow\downarrow\rangle + |\downarrow\uparrow\rangle)$, one can compute the expansion \eqref{defrho} of the density matrix for the general state \eqref{state}:%
\begin{eqnarray*}
 \rho &=& \left(\begin{tabular}{ccc} $a^2$ & $ab$ & $ac$  \\ $ab$ & $b^2$ & $bc$ \\ $ac$ & $bc$ & $c^2$ \end{tabular} \right) \\ &=& \frac 13 \bone \otimes \bone + \left(\frac{a^2+b^2}{2}-\frac 13\right)\left[\sigma_z \otimes \sigma_z - \frac 12(\sigma_x \otimes \sigma_x + \sigma_y\otimes \sigma_y)\right] + \frac{a^2-b^2}{2} \frac 12[\bone\otimes \sigma_z + \sigma_z \otimes \bone]\\
 && + \frac 12 ab[\sigma_x \otimes \sigma_x - \sigma_y \otimes \sigma_y] + \frac 12\frac{ac + bc}{\sqrt{2}}[\bone \otimes \sigma_x + \sigma_x \otimes \bone] + \frac 12\frac{ac - bc}{\sqrt{2}}\sigma_{(x}\otimes \sigma_{z)}\,.
\end{eqnarray*}
We see that in the state \eqref{state},
\be\label{p1p3}
p_1 = \frac{ac+bc}{\sqrt{2}}, \qquad p_3 = \frac{a^2-b^2}{2}\,.
\ee
As a consistency check, formula \eqref{entform} with the values \eqref{p1p3} for $p_1, p_3$ agrees with the EE from partial tracing out one of the spins in the state \eqref{state}. 

We claim that \eqref{entform} is the algebraic EE for the subalgebra generated by $\OO$ \eqref{defoo}. Note that $\OO^3 = (p_1^2+p_3^2)\OO$, so the subalgebra contains three independent elements, 
\be\label{subalg}
\AAA_0 = \{\bone \otimes \bone, \OO, \OO^2\}\,.
\ee To compute the algebraic EE, following the discussion in section \ref{s3}, we project the global state density matrix onto the operators of \eqref{subalg}.
The non-trivial step of this computation is to project the global density matrix along the plane in operator space spanned by $\bone \otimes \bone$ and 
\be\label{osquared}\begin{split}
\OO^2 = &\frac{2}{3}(p_1^2+p_3^2)\cdot \bone \otimes \bone + p_1p_3 \cdot \frac 12\sigma_{(x}\otimes \sigma_{z)} + \frac{p_1^2}{2}\cdot \frac 12(\sigma_x\otimes \sigma_x - \sigma_y \otimes \sigma_y) \\ & + \frac 1{2\sqrt{3}}(-p_1^2+2p_3^2)\cdot \frac{1}{\sqrt{3}}(\sigma_z \otimes \sigma_z - \frac 12(\sigma_x \otimes \sigma_x + \sigma_y \otimes \sigma_y))\,.
\end{split}\ee
Defining
$$\hat v = \frac{\sqrt{3}}{p_1^2+p_3^2}\left(\OO^2 - \frac 23(p_1^2+p_3^2)\cdot \bone \otimes \bone \right),  $$
 we find upon doing the projection that the algebraic EE of subalgebra \eqref{subalg} is the von Neumann entropy \eqref{defee} of 
 \be
 \rho_{\AAA_0} = \frac 13 \bone \otimes \bone +  p_1\cdot \frac 12(\bone \otimes \sigma_x + \sigma_x\otimes \bone) + p_3 \cdot  \frac 12(\bone \otimes \sigma_z + \sigma_z \otimes \bone) +\frac{1}{2\sqrt{3}} \hat v 
\ee
 in the 3d Hilbert space. 
 Computing it explicitly, we find agreement with \eqref{entform}, \eqref{p1p3}.
 

\end{appendix}

\newpage
\bibliographystyle{ssg.bst}
\bibliography{lgt}

\end{document}